\documentclass[aps,prb,twocolumn,superscriptaddress,longbibliography]{revtex4-2}
\usepackage[T1]{fontenc}
\usepackage[utf8]{inputenc}
\usepackage{graphicx}
\usepackage{subcaption}
\usepackage{amsmath}
\usepackage{amssymb}
\usepackage{braket}
\usepackage{bm}
\usepackage{bbm}
\usepackage{booktabs}
\usepackage{multirow}
\usepackage{microtype}
\usepackage{hhline}
\usepackage[colorlinks=true,linkcolor=blue,citecolor=blue]{hyperref}
\usepackage{orcidlink}
\usepackage{ragged2e}

\makeatletter
\long\def\@makecaption#1#2{%
  \vskip\abovecaptionskip
  \noindent\small\textbf{#1.}\hspace{0.3em}\justifying #2\par
  \vskip\belowcaptionskip}
\makeatother

\newcommand{\discussed}[1]{\textbf{#1}}

\begin{document}
\title{Thermodynamic Uncertainty Relations in Chaotic Andreev Billiards}

\author{I.\,R.\,A.\,C.~Lucena\,\orcidlink{0000-0002-3205-4228}}
\email{ivana.lucena@unesp.br}
\affiliation{Instituto de F\'isica Te\'orica,
  Universidade Estadual Paulista,
  S\~ao Paulo, SP, Brazil}
\affiliation{Laborat\'orio Nacional de Luz S\'incrotron,
  Centro Nacional de Pesquisa em Energia e Materiais,
  Campinas, SP, Brazil}

\author{T.\,J.\,A.~Mori\,\orcidlink{0000-0001-5340-3282}}
\affiliation{Laborat\'orio Nacional de Luz S\'incrotron,
  Centro Nacional de Pesquisa em Energia e Materiais,
  Campinas, SP, Brazil}

\author{M.\,M.~Soares\,\orcidlink{0000-0001-8128-7901}}
\affiliation{Departamento de F\'isica,
  Universidade Federal da Para\'iba,
  Jo\~ao Pessoa, PB, Brazil}

\author{D.~Bazeia\,\orcidlink{0000-0003-1335-3705}}
\affiliation{Departamento de F\'isica,
  Universidade Federal da Para\'iba,
  Jo\~ao Pessoa, PB, Brazil}

\author{A.\,R.~Rocha\,\orcidlink{0000-0001-8874-6947}}
\affiliation{Instituto de F\'isica Te\'orica,
  Universidade Estadual Paulista,
  S\~ao Paulo, SP, Brazil}

\date{\today}
\begin{abstract}
We investigate how particle-hole symmetry, quantum interference, and tunnel barriers shape thermodynamic uncertainty relations in chaotic Andreev billiards. Using random-matrix theory and the Mahaux--Weidenm\"uller scattering approach, we study charge conductance and shot noise across the four Altland--Zirnbauer symmetry classes, from the single-channel extreme quantum limit to the multichannel semiclassical regime and from ideal to opaque contacts. We characterize thermodynamic precision through two complementary ensemble observables: either by averaging the sample-resolved noise-to-conductance ratio, or from the separately averaging noise and conductance. Their pronounced discrepancy in the extreme quantum regime reveals the non-self-averaging character of mesoscopic transport and persists over a broad range of barrier transparencies. For ideal contacts, the first provides a sensitive fingerprint of the Altland--Zirnbauer symmetry class. In the opaque regime, this hierarchy changes. Despite these strong symmetry- and barrier-dependent effects, the standard thermodynamic uncertainty relation remains satisfied throughout all regimes investigated. Our results establish thermodynamic uncertainty as a symmetry-sensitive probe of universal transport in chaotic normal--superconducting systems.

\end{abstract}

\maketitle

\section{Introduction}
\label{sec:intro}

In his seminal work on irreversibility in computation, Rolf Landauer raised the fundamental question of whether information processing is ultimately constrained by physical principles~\cite{5392446}. At the quantum scale, addressing this question requires understanding how thermodynamic constraints coexist with fluctuations, coherence, and quantum transport. This issue is particularly relevant for mesoscopic devices, where charge is transferred by a small number of carriers and fluctuations become an intrinsic component of transport rather than a negligible correction~\cite{koski2014experimental,RevModPhys.85.1421,gustavsson2006counting}.

The thermodynamic cost of achieving precise transport is quantified by the thermodynamic uncertainty relations (TUR), which establish a trade-off between current fluctuations and entropy production~\cite{barato2015thermodynamic,horowitz2020thermodynamic,gingrich2017fundamental,timpanaro2019thermodynamic,timpanaro2025quantum}. For charge transport, the conventional TUR can be expressed as
\begin{equation}\label{eq:1}
\mathcal{Q}\equiv\frac{\langle\sigma\rangle S}{I^2k_B} \geq 2,
\end{equation}
where $I$ is the mean current, $S$ is the current noise, and $\langle\sigma\rangle$ is the average entropy production rate~\cite{PhysRevE.93.052145,PhysRevB.101.195423}. Originally derived for classical Markovian dynamics, the TUR have, subsequently, been extended to phase-coherent conductors~\cite{PhysRevLett.116.120601,6nww-8wcp,tesser2024out,potanina2021thermodynamic}, where quantum interference can substantially modify the relation between fluctuations, dissipation, and transport precision.

Chaotic mesoscopic conductors provide a particularly useful setting for investigating these effects. Quantum coherence produces characteristic interference phenomena such as weak localization (WL) and universal conductance fluctuations~\cite{ramos2008quantum,PhysRevB.84.035453,PhysRevLett.107.176807,heinzel2008mesoscopic}. Despite the complexity of the underlying chaotic dynamics, transport becomes universal, i.e., the statistical properties of observables are determined primarily by global symmetries rather than microscopic details. Random matrix theory (RMT), therefore, provides a natural framework for describing transport in these systems~\cite{PhysRevLett.93.014103,RevModPhys.69.731,mehta2004random}.

Hybrid normal--superconducting (NS) structures enrich this scenario through Andreev reflection, in which electron and hole degrees of freedom become coherently coupled~\cite{de2010hybrid,burkard2020superconductor,RevModPhys.93.041001}. Besides their relevance to quantum-information architectures~\cite{benito2020hybrid,clarke2008superconducting,PhysRevX.5.041038} and quantum thermodynamic devices~\cite{tabatabaei2022nonlocal,benenti2017fundamental,whitney2019quantum,blasi2023hybrid}, NS systems offer a distinctive setting for studying fluctuation constraints because Andreev processes strongly modify both charge-transfer statistics and current noise. At the symmetry level, the intrinsic particle--hole symmetry (PHS) of the Bogoliubov--de Gennes Hamiltonian extends the conventional Wigner--Dyson classification. Together with time-reversal symmetry (TRS) and spin-rotation symmetry (SRS), it gives rise to the four Altland--Zirnbauer (AZ) symmetry classes relevant to NS systems~\cite{Altland1997,Beenakker2015}. Consequently, chaotic Andreev billiards provide a natural platform for investigating how superconducting correlations and fundamental symmetries shape thermodynamic precision.

In this work, we employ RMT to investigate TUR in open chaotic Andreev billiards across all four Altland--Zirnbauer symmetry classes with PHS. The scattering problem is implemented numerically through a Monte Carlo realization of the Mahaux--Weidenm\"{u}ller Hamiltonian. We explore the crossover from the extreme quantum limit, characterized by a single open transport channel, to the semiclassical multichannel regime, as well as the full range of barrier transparencies from ideal contacts to the opaque tunneling limit. A central feature of the NS problem is that Andreev reflection maps the normal-state transmission eigenvalues nonlinearly onto the corresponding Andreev transmission eigenvalues. This transformation reshapes the conductance and noise statistics and, consequently, the TUR observable itself.

Our main findings are threefold. First, for ideal contacts, the TUR observable provides a sensitive fingerprint of the full Altland--Zirnbauer symmetry class, encoding the combined effects of particle--hole, time-reversal, and spin-rotation symmetries, in direct analogy with the chirality fingerprint previously identified in Dirac billiards. Second, unlike in Wigner--Dyson and chiral systems, the TRS-driven quantum-interference correction does not provide the dominant contribution to the TUR in Andreev billiards. This does not result from suppressed interference; rather, Andreev reflection enhances the semiclassical contribution through its nonlinear modification of the transmission-eigenvalue statistics, thereby reducing the relative weight of the interference correction. Third, in the tunneling regime, the TUR becomes even more sensitive to the underlying symmetry class than for ideal contacts, a behavior that we characterize through two complementary TUR observables. Albeit the bound $\mathcal{Q} \geq 2$ remains satisfied.

The remainder of this paper is organized as follows. In Sec.~\ref{sec:billiard}, we introduce the RMT description of the chaotic Andreev billiard and the corresponding theoretical framework. In Sec.~\ref{sec:transport}, we analyze the ensemble-averaged conductance and shot-noise power for the four AZ symmetry classes as functions of the barrier transparency. In Sec.~\ref{sec:TUR}, we investigate the thermodynamic uncertainty relations and discuss the roles of quantum interference and symmetry, followed by our conclusions.

\section{Open Chaotic Andreev Billiard}
\label{sec:billiard}

Our starting point is to establish the connection between RMT and transport observables, providing a framework and numerical results for the study of TUR in the universal regime.

\subsection{Model and approach}
\label{sec:A}
\begin{figure}[t]
\centering
\includegraphics[width=1.0\linewidth]{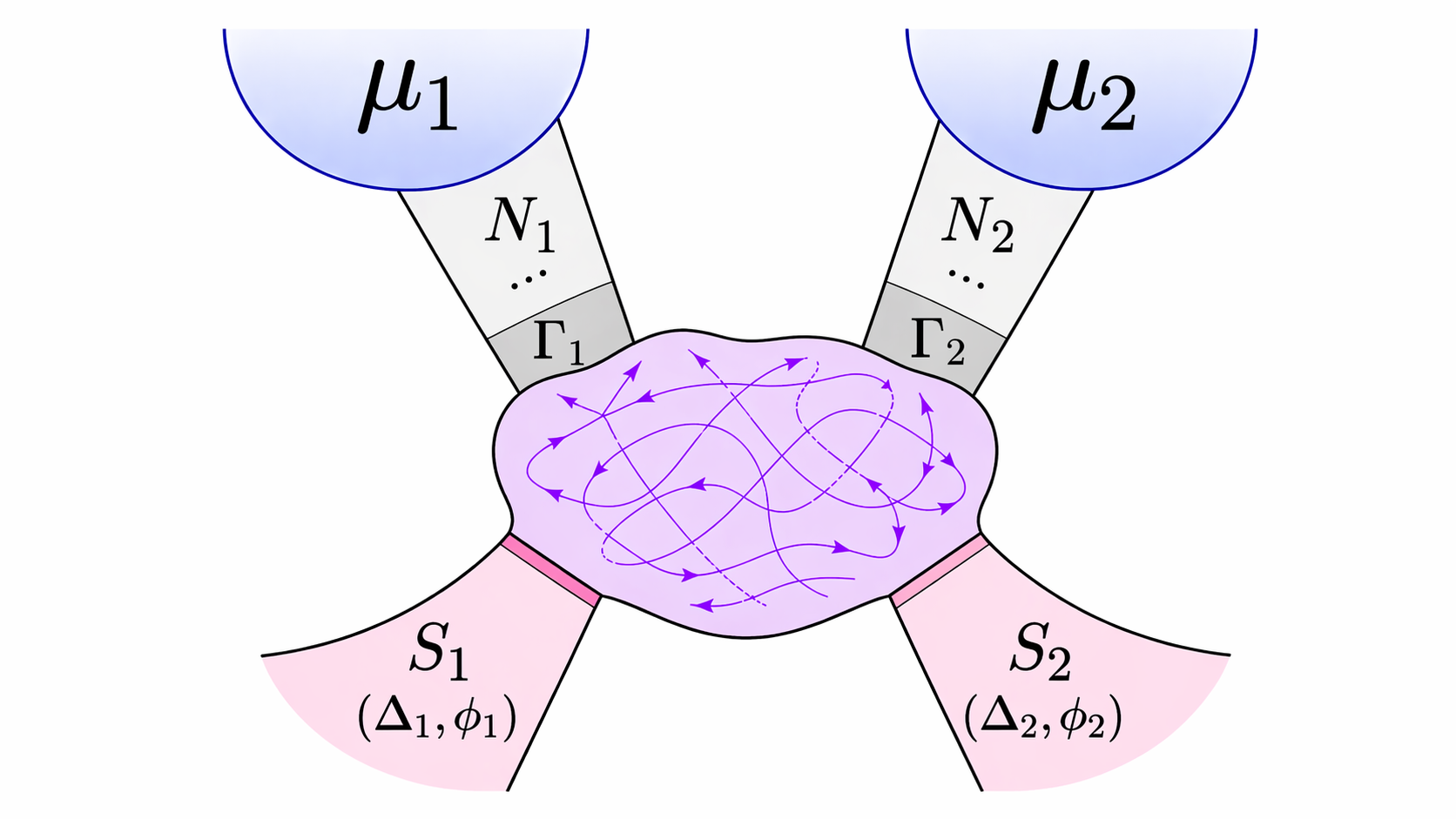}
\caption{Schematic view of the chaotic Andreev billiard in the
universal regime. The mesoscopic cavity is coupled to two
normal-metal electron reservoirs at chemical potentials $\mu_1$ and
$\mu_2$ via leads supporting $N_1$ and $N_2$ open channels, and to superconducting contacts, $S_{1}$ and $S_{2}$. In the ballistic regime,
electrons undergo multiple coherent reflections, leading to chaotic scattering.}
\label{Fig1}
\end{figure}

We deal with an open chaotic Andreev billiard (AB), as shown in Fig. \ref{Fig1}. The system consists of a phase-coherent mesoscopic cavity embedded in a two-dimensional electron gas (2DEG) connected to superconducting electrodes~\cite{Beenakker2015}. The cavity is connected to two normal leads supporting $N_1$ and $N_2$ propagating 
modes, with chemical potentials $\mu_1$ and $\mu_2 = \mu_1 - eV$, and to two 
$s$-wave superconductors, $S_{1}$ and $S_{2}$,  with pairing gap $\Delta_{1} = \Delta_{2}=\Delta$. The superconductors are biased 
with a phase difference of $\phi_1 - \phi_2 = \pi$, which suppresses the proximity-induced minigap in 
the normal region, ensuring gapless quasiparticle excitations at the Fermi level~\cite{dahlhaus2010random}. This condition is essential for ergodic exploration 
of phase space and, consequently, for the onset of fully chaotic scattering in the 
AB. Since phase differences of $\pi$ and $-\pi$ are physically equivalent, time-reversal symmetry (TRS) is preserved.

Electronic transport is governed by Andreev reflection at the normal-superconductor interfaces, i.e., an electron with excitation energy $E<\Delta$ is backscattered as a hole, while a Cooper pair is absorbed into the condensate. This mechanism leads to a nonlinear relation between conductance and transmission eigenvalues~\cite{beenakker2005andreev}. In the presence of superconducting order, for spin-singlet $s$-wave pairing, the mean-field Hamiltonian can be written in the Bogoliubov-de Gennes (BdG) form~\cite{Altland1997,Beenakker2015}
\begin{equation}
    H = \hat{\Psi}^\dagger \mathcal{H} \hat{\Psi},
    \qquad
    \mathcal{H} =
    \begin{pmatrix}
        H_0 - E_{F} & -i\hat{\sigma}_y \Delta \\
        i\hat{\sigma}_y \Delta^* & E_F - H_0^*
    \end{pmatrix},
    \label{eq:BdG}
\end{equation}
where $\hat{\Psi}=
(\hat{\psi}_\uparrow, \hat{\psi}_\downarrow, \hat{\psi}^\dagger_\uparrow,
\hat{\psi}^\dagger_\downarrow)^{T}$ is a four-component Nambu spinor, $H_0$ is the single-particle Hamiltonian, and $\Delta$ is the (complex) superconducting order parameter. The off-diagonal blocks describe pairing between electrons of opposite spin, and $\hat{\sigma}_{y}$ is a Pauli matrix, which acts on spin degrees of freedom.  

It is easy to see that the BdG Hamiltonian satisfies the particle-hole symmetry,
\begin{equation}
    \mathcal{H} = -\,\mathcal{C}\,\mathcal{H}\,\mathcal{C}^{-1}
                = -\,\tau_x\,\mathcal{H}^*\,\tau_x,
    \label{eq:ph-sym}
\end{equation}
where $\mathcal{C} = \tau_x\,\mathcal{K}$, with $\tau_x$ being Pauli matrix acting in Nambu (particle-hole) space and $\mathcal{K}$
denotes the operator of complex conjugation, which is anti-unitary and squares to $+1$. The $\mathcal{H}$ matrix has dimension $2M\times 2M$, where
$M$ is the number of electronic degrees of freedom in the normal region. The Hamiltonian model for the scattering matrix, $\mathcal{S}$, can be written as~\cite{mahaux1971shell,mahaux1969fine,mahaux1968comparison}
\begin{equation}\label{MWscattering}
\mathcal{S} = \textbf{1} - 2\pi i\mathcal{W}^{\dagger}(E - \mathcal{H}+ i\pi \mathcal{W}\mathcal{W}^{\dagger})^{-1}\mathcal{W},
\end{equation}
where $E$ denotes the quasiparticle excitation energy and $\mathcal{W}$ represents all combinations (interactions) of the dot's resonances coupled to the open channels of the terminals.  The matrix $\mathcal{S}$ has dimensions $N_{T}\times N_{T}$, where $N_{T}=N_{1}+N_{2}$ is the total number of open channels in the terminals that are connected to the quantum dot. Furthermore, it is convenient to represent $\mathcal{S}$ as a function of transmission, $t$, and reflection, $r$, blocks as 
\begin{equation}
\mathcal{S} =
\begin{pmatrix}
    r_{N_{1} \times N_{1}} & t'_{N_{1} \times N_2} \\
    t_{N_2 \times N_1} & r'_{N_2 \times N_2}
\end{pmatrix}
\label{eq:Smatrix},
\end{equation}
where $t$ ($t'$) and $r$ ($r'$) are the transmission and reflection
sub-matrices for the modes incident from lead 1 (lead 2).

\subsection{Electronic Transport Properties}
\label{sec:transport}

Since the TUR, whose validity we want to address, can be expressed as a ratio between the shot-noise 
power and the electrical conductance, both quantities must be computed for each 
relevant symmetry class. However, analytical results for transport quantities in 
AB are generally restricted to the ideal-contact limit~\cite{10.1063/1.531667}, which excludes the experimentally relevant case of opaque contacts. To overcome this limitation, we develop a numerical simulation based on RMT~\cite{mehta2004random}, assuming that both the dephasing time $\tau_\phi$ and the electronic dwell time $\tau_D$ satisfy $\{\tau_\phi,\, \tau_D \gg \tau_E\}$, where $\tau_E$ is the Ehrenfest time. This condition ensures that the system operates in the universal RMT regime, in which results are independent of microscopic details.

The symmetry classification of the problem follows from the particle-hole symmetry 
of the Bogoliubov--de~Gennes Hamiltonian, which naturally extends the standard 
Wigner--Dyson classification and gives rise to four additional symmetry 
classes~\cite{Altland1997}. Depending on the presence or absence of TRS and SRS, each class is characterized by a index $\beta_{E}$ and a second symmetry index $\gamma$. 
Table~\ref{table:AZclassification} summarizes the AZ symmetry classes of $\mathcal{S}$ and the corresponding parameters characterizing them. For each class, we generate ensembles of scattering matrices spanning both the opaque and ideal coupling regimes, from which the TUR ratio is extracted and systematically compared across coupling strengths.
\begin{table}[h]
\caption{Classification of the Altland--Zirnbauer scattering matrix ensembles for superconducting systems.}
\label{table:AZclassification}
\begin{ruledtabular}
\begin{tabular}{lcccccc}
Ensemble & Class & TRS & SRS & $S$-space & $\beta_{E}$ & $\gamma$ \\
\hline
CRE   & D    & $\times$     & $\times$     & Orthogonal           & 1 & $-1$ \\
T-CRE & DIII & $\checkmark$ & $\times$     & Orth. self-dual      & 2 & $-1$ \\
CQE   & C    & $\times$     & $\checkmark$ & Symplectic           & 4 & $2$  \\
T-CQE & CI   & $\checkmark$ & $\checkmark$ & Sympl. symmetric     & 2 & $1$  \\
\end{tabular}
\end{ruledtabular}
\end{table}

We use Eq.(\ref{MWscattering}) to construct random scattering matrices with PHS. The RMT establishes that the entries of $\mathcal{H}$ matrix have Gaussian distribution, 
\begin{equation}
    P(\mathcal{H}) \propto \exp\!\left( -\frac{\pi^2 \beta_{E}}{8 M \tilde{\Delta}^2} \,\mathrm{Tr}\, \mathcal{H}^2 \right)
\end{equation}
where $\tilde{\Delta}$ is the mean level spacing of $\mathcal{H}$ in the bulk spectrum. In order to forbid direct processes, i.e., processes in which electrons do not traverse the quantum dot before scattering, the deterministic matrix $\mathcal{W}$ must satisfy the orthogonality condition~\cite{verbaarschot1985grassmann}
\begin{equation}
\mathcal{W}_{\alpha}\mathcal{W}_{\beta}^{\dagger} =
\lambda_{\alpha}\frac{M\tilde{\Delta}}{\pi^2}\delta_{\alpha,\beta},
\label{eq:ortho}
\end{equation}
in which $\lambda_{\alpha}$ is a diagonal matrix containing the tunneling rates $\Gamma_{\alpha,a} \in [0,1]$ of the $a$ channel supported by the lead $\alpha$ via
$\Gamma_{\alpha,a} = \mathrm{sech}^2[-\ln(\lambda_{\alpha a})/2]$. 
Throughout this work we take equal tunneling probabilities,
$\Gamma \equiv \Gamma_{\alpha,a}$ for all $(\alpha,a)$, interpolating between the ideal ballistic ($\Gamma = 1$) and opaque tunneling ($\Gamma \to 0$) regimes. To ensure the chaotic regime and consequently universality of observables, the number of resonances inside the chaotic billiard is large ($M\gg N_{T}$).

\begin{figure}[h]
\centering
\includegraphics[width=1.0\linewidth]{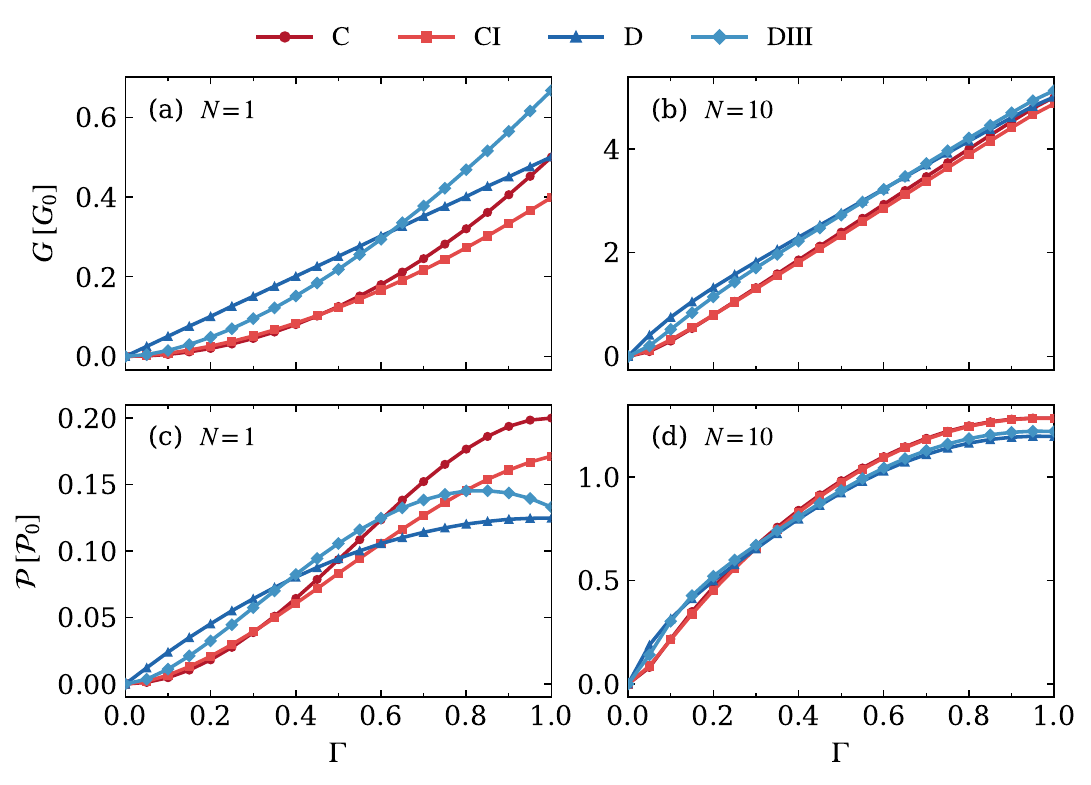}
\caption{Ensemble averages of the charge conductance $G$
(upper panels) and shot-noise power $\mathcal{P}$ (lower
panels) as a function of barrier transparency $\Gamma$ for all four
Altland--Zirnbauer classes, for $N=1$ (left) and $N=10$ (right).}
\label{fig:WL_Wigner}
\end{figure}

Following the Landauer--B\"uttiker formalism~\cite{imry1999conductance,datta1997electronic}, the conductance, $G$, is obtained from the transmission matrix block of 
$\mathcal{S}$, Eq.(\ref{eq:Smatrix}), as
\begin{equation}
    G = G_0\,\mathrm{Tr}\!\left(tt^{\dagger}\right),
\end{equation}
where we define the quantum of conductance $G_0=2e^2/h$. The ensemble average, $[G]$, is obtained over Hamiltonian realizations. In the universal regime, the averages are isospectral, i.e., we can execute them by setting, for simplicity, $E=0$. In addition, we perform a numerical simulation of the average shot-noise power, $[\mathcal{P}]$, which is the amplitude of the long time current correlation and characterizes the discreteness of the charge transport process. At zero temperature, the shot-noise power is given by 
\begin{equation}
    \mathcal{P} = \frac{4e^3|V|}{h}\textbf{Tr}[tt^{\dagger}(1-tt^{\dagger})].
\end{equation}
where $V$ is the bias voltage and we define $\mathcal{P}_{0}=4e^{3}|V/h|$. Fig.(\ref{fig:WL_Wigner}) shows the ensemble-averaged conductance
$G$ and shot-noise power $\mathcal{P}$ as a function of
the barrier transparency $\Gamma$ for all four AZ classes, in both
the $N=1$ and $N=10$ regimes. The average values of G and $\mathcal{P}$ were obtained from $10^{6}$ realizations, for each class. For class D, we use random Hamiltonians with dimensions $100\times 100$. For class DIII, the Hamiltonians have dimensions $200\times 200$ due to the spin degrees of freedom. Both classes C and CI were generated by random Hamiltonians with dimensions $200\times 200$ due to the degrees of freedom of the Bogoliubov quasiparticles.

The conductance and shot-noise power for all four classes are shown in Fig.~\ref{fig:WL_Wigner}(a) and \ref{fig:WL_Wigner}(c) for the extreme quantum regime ($N=1$). Although the conductance increases monotonically with the barrier transparency $\Gamma$, it displays a markedly non-Ohmic behavior that is quantitatively distinct across the classes. This distinction arises from high-order quantum interference corrections (QIC) generated by multiple coherent scattering events within the mesoscopic cavity, whose structure is governed by the underlying symmetry class. To isolate this TRS-driven QIC from our numerical results, we employ the subtraction procedure of~\cite{lucena2023thermodynamic}. It consists of taking the transport observables of the corresponding TRS-broken classes (D and C), which share 
the same particle-hole symmetry structure but lack the 
cooperon modes generated by TRS, as the semiclassical 
baseline. Subtracting them from the DIII and CI results 
removes the PHS common quantum contributions present in all four AZ classes, thereby isolating the purely 
TRS-driven interference component. 

\begin{figure}[t]
\centering
\includegraphics[width=1.0\linewidth]{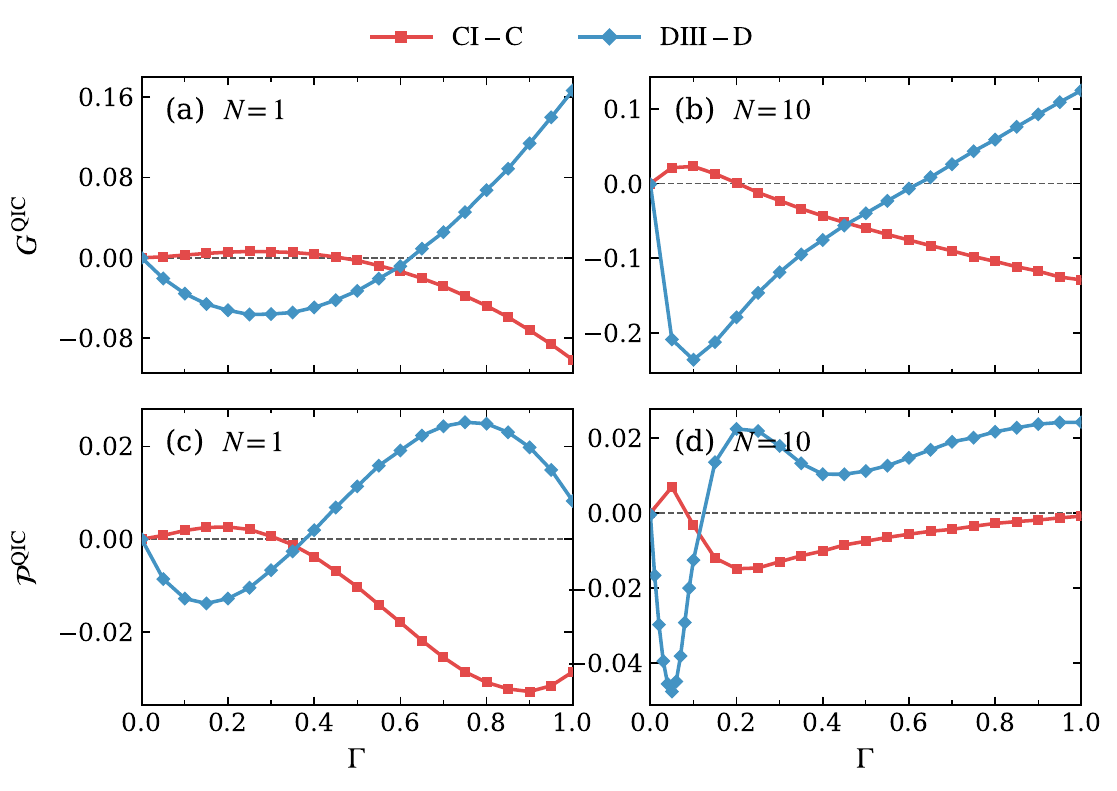}
\caption{Weak-localization correction $G^{\mathrm{QIC}}$ to the
conductance (upper panels) and the analogous correction
$\mathcal{P}^{\mathrm{QIC}}$ to the shot-noise power (lower panels)
as a function of $\Gamma$ for all four Altland--Zirnbauer classes,
for $N=1$ (left) and $N=10$ (right).}
\label{FigQIC}
\end{figure}

The quantum interference sector, depicted in Fig.~\ref{FigQIC}, reveals a pronounced barrier-induced suppression-amplification transition for both the conductance and the shot-noise power, i.e., as $\Gamma$ is varied, the QICs change sign, passing from regions of weak antilocalization (positive sign) to weak localization (negative sign). This non-monotonic behavior 
reflects how time-reversal symmetry governs phase coherence of electronic paths inside the billiard. 

In the semiclassical regime ($N=10$), shown in Figs.~\ref{fig:WL_Wigner}(b) and~\ref{fig:WL_Wigner}(d), all conductance curves recover a nearly Ohmic behavior, while the average shot-noise power exhibits behavior characteristic of a Schr\"odinger billiard. However, the presence of quantum interference corrections (QIC) in the shot-noise power remains evident in Andreev billiards even for the semiclassical regime. As illustrated in Fig.~\ref{FigQIC}, the QIC undergoes a sign change as a function of $\Gamma$, revealing a crossover between transport regimes. Since spin-rotation symmetry is preserved (broken) in class CI (DIII), we observe that its invariance is encoded in the QIC.

\section{ Thermodynamic Uncertainty Relations in Andreev Billiards}
\label{sec:TUR}

Conductance and shot-noise power measurements provide powerful tools for probing charge transport in mesoscopic systems. However, in the context of energy-converting devices, where the goal is to generate currents of significant magnitude with controlled precision, it is equally important to assess the thermodynamic cost of such precision. This can be quantified by the thermodynamic uncertainty relations (TUR), which relate the steady-state charge current $\langle I\rangle$, its fluctuations $\langle\langle I^2\rangle\rangle = \langle I^2\rangle -\langle I\rangle^2$, and the average entropy production due to Joule heating, given by $\langle \sigma\rangle =\langle I\rangle V/T $. Thus, 
\begin{equation}
\frac{\langle\langle I^2\rangle\rangle}{\langle I\rangle^2} \geq \frac{2k_{B}}{\langle \sigma\rangle} \xrightarrow{\textbf{Joule's Law}} \frac{V}{k_{B}T}\frac{\langle\langle I^2\rangle\rangle}{\langle I\rangle} \geq 2
\end{equation}
where $T$ is the temperature of the electronic reservoirs. Using Eq.\ref{eq:1}, it is convenient to redefine the combination $\mathcal{Q}\equiv \frac{V}{k_{B}T}\frac{\langle\langle I^2\rangle\rangle}{\langle I\rangle}$, which serves as the TUR observable. If the system satisfies $\mathcal{Q}\geq 2$, the TUR holds. In what follows, we show that universal symmetry properties, quantum chaos and quantum interference play a decisive role in determining whether this bound is satisfied.

Within the Landauer--B\"uttiker formalism in the linear regime and steady sate~\cite{lucena2023thermodynamic}, the TUR can be naturally expressed in terms of the ratio between the shot-noise power and the conductance. In the RMT framework, however, a subtlety arises when performing the ensemble average because the ratio of two averaged quantities is not, in the quantum extreme limit, equal to the average of their ratio. This ambiguity motivates the introduction of two distinct observables. The first, which we call the $\mathcal{Q}_{\mathrm{R}}$, is obtained by computing the ratio for each sample and then averaging over the ensemble,
\begin{equation}\label{tur_R}
    \mathcal{Q}_{\mathrm{R}} = \frac{V}{k_{B}T}\left[\frac{\mathcal{P}}{G}\right],
\end{equation}
where $[\dots]$ denotes an ensemble average. The second observable, which we call $\mathcal{Q}_{\mathrm{MS}}$, is instead constructed by averaging the noise and conductance separately before taking their ratio, 
\begin{equation}\label{tur_MS}
\mathcal{Q}_{\mathrm{MS}} = \frac{V}{k_{B}T}\frac{[\mathcal{P}]}{[ G]}.
\end{equation}
These two quantities coincide only when sample-to-sample fluctuations are negligible, and their comparison therefore provides a direct probe of mesoscopic fluctuations effects on the TUR. Throughout this section, we restrict our analysis to the TRS-preserving classes DIII and CI, under a fixed bias $\delta\mu = 0.05\mathrm{V}$ and inverse temperature $(k_B T)^{-1} \approx 2000\mathrm{eV}^{-1}$.
\begin{figure}[!h]
\centering
\includegraphics[width=1.0\linewidth]{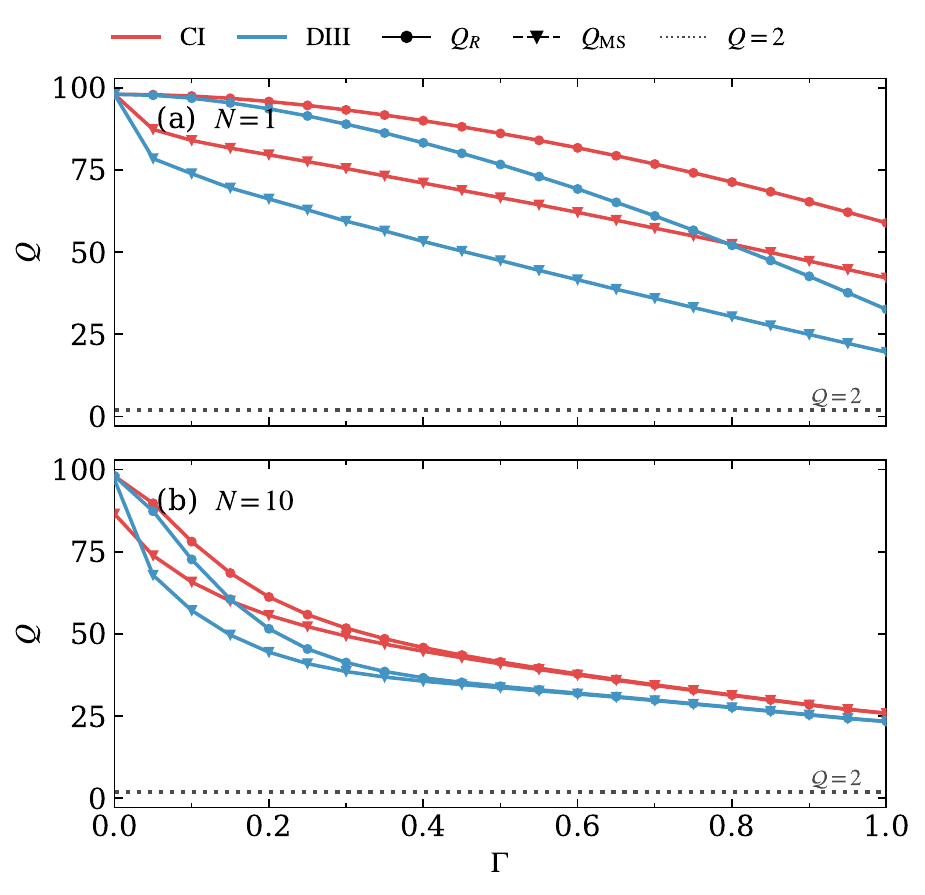}
\caption{Total factors $Q_{\mathrm{R}}$ and
$Q_{\mathrm{MS}}$ for the DIII (T-CRE) and CI (T-CQE)
Altland--Zirnbauer symmetry classes as a function of $\Gamma$.
Both observables satisfy the TUR bound $\mathcal{Q} \geq 2$ for all values of $\Gamma$ investigated. Parameters: $\mu = 0.05$\,V and $(k_BT)^{-1} \approx 2000$\,eV$^{-1}$.}
\label{FigQfull}
\end{figure}

In Fig. $\ref{FigQfull}$ we display the signal-to-noise ratio, $\mathcal{Q}_{\mathrm{R}}$, and $\mathcal{Q}_{\mathrm{MS}}$ as a function of $\Gamma$ for the Andreev chaotic billiard, in two representative regimes, the extreme quantum regime ($N=1$), shown in Fig. \ref{FigQfull}(a), and the semiclassical regime ($N=10$), shown in Fig. \ref{FigQfull}(b).
\begin{figure}[h]
\centering
\includegraphics[width=1.0\linewidth]{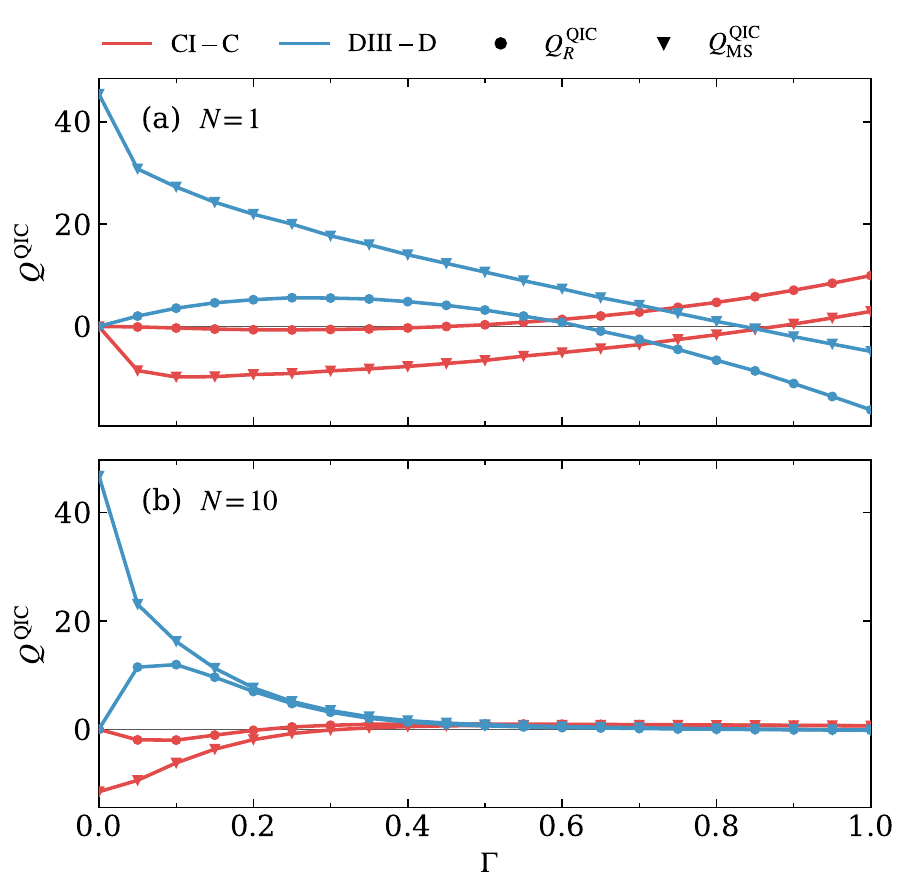}
\caption{Quantum interference correction $\mathcal{Q}^{\mathrm{QIC}}$
to the TUR-R and TUR-MS for the DIII (T-CRE) and CI (T-CQE)
Altland--Zirnbauer symmetry classes, as a function of barrier
transparency $\Gamma$, for $N=1$ (upper) and $N=10$ (lower).
Parameters: $\mu = 0.05$\,V and $(k_BT)^{-1} \approx 2000$\,eV$^{-1}$.}
\label{FigQ}
\end{figure}
In both regimes, the TUR observables differ as $\Gamma\rightarrow 0$, reflecting the suppression of transport through the barrier. More importantly, $\mathcal{Q}_{\mathrm{R}}$ and $\mathcal{Q}_{\mathrm{MS}}$ yield distinct values across a wide range of $\Gamma$, including the limit $\Gamma \rightarrow 1$. This discrepancy is particularly pronounced in the extreme quantum regime, but persists over large intervals of barrier strength even in the semiclassical case. Such a disagreement between the two observables signals a breakdown of the law of large numbers and the central limit theorem, an effect previously observed for Wigner-Dyson and Chiral symmetry classes \cite{lucena2023thermodynamic}, and here extended to the Andreev (superconductor) setting.

Another important feature is that the TUR bound, $\mathcal{Q} \geq 2$, is consistently satisfied for both observables across all symmetry classes and for the entire range of $\Gamma$ considered. This result demonstrates the absence of TUR violations in chaotic Andreev billiards, indicating that charge transport remains fully compatible with thermodynamic uncertainty constraints in the universal regime. The quantitative comparison between $\mathcal{Q}_{\mathrm{R}}$ and $\mathcal{Q}_{\mathrm{MS}}$ is summarized
in Table~\ref{table_II}, along with the results
for the Wigner-Dyson and Chiral classes as a reference \cite{lucena2023thermodynamic}.
\begin{table*}[t]
\centering
\caption{Comparison between ideal ($\Gamma = 1.0$) and opaque ($\Gamma = 0.1$) regimes.
We report the quantum interference contribution $\mathcal{Q}^{\mathrm{QIC}}$,
the total TUR value $\mathcal{Q}$, and the relative contribution
$\mathcal{Q}^{\mathrm{QIC}}/\mathcal{Q}$ for $N=1$ and $N=10$,
across all symmetry classes. Boldface indicates values explicitly discussed
in the main text.}
\label{table_II}

\begin{ruledtabular}
\begin{tabular}{lll
                cc cc cc
                cc cc cc}

& Class & Ensemble
& \multicolumn{6}{c}{$\Gamma=1.0$}
& \multicolumn{6}{c}{$\Gamma=0.1$}
\\

\cline{4-9}
\cline{10-15}

& &
& \multicolumn{2}{c}{$\mathcal{Q}^{\mathrm{QIC}}$}
& \multicolumn{2}{c}{$\mathcal{Q}$}
& \multicolumn{2}{c}{$\mathcal{Q}^{\mathrm{QIC}}/\mathcal{Q}$ (\%)}
& \multicolumn{2}{c}{$\mathcal{Q}^{\mathrm{QIC}}$}
& \multicolumn{2}{c}{$\mathcal{Q}$}
& \multicolumn{2}{c}{$\mathcal{Q}^{\mathrm{QIC}}/\mathcal{Q}$ (\%)}
\\

& &
& $N=1$ & $N=10$
& $N=1$ & $N=10$
& $N=1$ & $N=10$
& $N=1$ & $N=10$
& $N=1$ & $N=10$
& $N=1$ & $N=10$
\\

\hline

\multirow{6}{*}{$\mathcal{Q}_{\mathrm{R}}$}

& WD & Orth.
& 58.90 & 1.28
& 65.38 & 25.91
& 90.0 & 4.94
& 83.0 & 2.39
& 95.31 & 67.33
& 87.08 & 3.54
\\

& & Sympl.
& 26.80 & $-$0.71
& 32.66 & 23.91
& 82.0 & 2.96
& 81.7 & $-$2.06
& 93.76 & 63.0
& 87.13 & 3.27
\\

& Ch & Orth.
& 34.20 & 0.04
& 38.74 & 24.59
& 88.2 & 0.1
& 22.0 & $-$9.30
& 87.1 & 47.9
& 25.2 & 19.41
\\

& & Sympl.
& 27.8 & $-$0.013
& 33.10 & 24.53
& 83.9 & 0.05
& 30.9 & 7.21
& 95.6 & 64.37
& 32.33 & 11.2
\\

& AZ & CI
& 9.87 & 0.66
& 58.60 & 25.11
& \discussed{16.84} & 2.62
& $-$0.5 & $-$2.05
& 97.0 & 77.5
& \discussed{0.51} & 2.64
\\

& & DIII
& $-$16.44 & $-$0.18
& 32.2 & 23.32
& \discussed{51.05} & 0.77
& 3.6 & 11.72
& 96.0 & 72.1
& \discussed{3.75} & \discussed{16.25}
\\

\hline

\multirow{6}{*}{$\mathcal{Q}_{\mathrm{MS}}$}

& WD & Orth.
& 7.3 & 1.20
& 39.26 & 25.78
& 18.5 & 4.65
& 2.90 & 1.52
& 61.43 & 50.28
& 4.72 & 3.02
\\

& & Sympl.
& $-$7.9 & $-$0.66
& 24.81 & 23.89
& 31.84 & 2.76
& $-$3.3 & $-$0.92
& 55.35 & 47.77
& 5.96 & 1.92
\\

& Ch & Orth.
& $-$0.58 & 0.01
& 27.45 & 24.53
& 2.11 & \discussed{0.04}
& $-$26.0 & $-$9.30
& 48.2 & 41.61
& 53.94 & \discussed{22.35}
\\

& & Sympl.
& 1.34 & 0.002
& 29.33 & 24.52
& 4.56 & 0.08
& 18.7 & 9.21
& 92.8 & 60.2
& 20.15 & 15.3
\\

& AZ & CI
& 2.86 & 0.60
& 41.9 & 25.5
& \discussed{6.82} & 2.35
& $-$10.1 & $-$6.31
& 83.72 & 65.2
& \discussed{12.06} & 9.67
\\

& & DIII
& $-$5.08 & $-$0.10
& 18.9 & 22.5
& \discussed{26.87} & 0.44
& 27.4 & 16.16
& 73.4 & 56.64
& \discussed{37.32} & \discussed{28.53}
\\

\end{tabular}
\end{ruledtabular}
\end{table*}

These findings can be placed in the broader context of recent investigations on TUR behavior in normal-superconducting hybrid systems. In Ref.~\cite{manzano2023quantum}, Manzano \textit{et al.} demonstrated TUR violations in a single quantum dot coupled to a normal reservoir and a superconducting electrode in the subgap regime, where they arise from quantum coherence induced by the superconducting proximity effect and occurring preferentially at high efficiency and near-maximum power output. More recently, results of Ref. \cite{ohnmacht2025thermodynamic} showed that TUR breaking in NS and SS quantum point contacts with energy-independent transmission requires high channel transmissions ($\tau \geq 0.91$) and is driven by the coexistence of quasiparticle tunneling and Andreev reflection processes, with SS junctions exhibiting violations roughly one order of magnitude larger than NS ones. In contrast, here we investigated a distinct and complementary regime within this landscape. Rather than a few-mode junction, we studied an open chaotic Andreev billiard in the universal RMT regime, characterized by a large number of internal resonances ($M \gg N_{T}$) and covering the full crossover from the extreme quantum ($N=1$) to the semiclassical ($N=10$) limit.

The TUR receives distinct contributions: one is denoted as the semiclassical component common to all symmetry classes sharing the same particle-hole structure, and the other, $\mathcal{Q}^{\mathrm{QIC}}$, accounts for quantum interference corrections driven exclusively by time-reversal symmetry. 

A central result of our analysis is that, in contrast to the Schr\"{o}dinger and Dirac billiards, the QIC is \emph{not} the dominant contribution to
the TUR in Andreev billiards. This behavior does not reflect a suppression of the interference term itself, but rather an amplification of the semiclassical baseline. The Andreev reflection modifies the nonlinear relation between  conductance, shot-noise power, and transmission eigenvalues,  enhancing the TUR well above the quantum bound $\mathcal{Q}\geq 2$ 
already at the semiclassical level. As a consequence, the 
relative contribution $\mathcal{Q}^{\mathrm{QIC}}/\mathcal{Q}$ is reduced compared to normal-conducting billiards. 

In the extreme quantum regime ($N=1$), sample-to-sample fluctuations remain dominant, leading to a pronounced discrepancy between the two TUR observables, $\mathrm{TUR}\text{-}\mathrm{R}$ and $\mathrm{TUR}\text{-}\mathrm{MS}$. In this regime, both observables exhibit a sign change of $\mathcal{Q}^{\mathrm{QIC}}$ as a function of $\Gamma$, with a positive-to-negative transition for class DIII and a negative-to-positive transition for class CI, as shown in Fig.~\ref{FigQ}(a). Such transitions reflect the interplay between the TRS-driven cooperon modes and barrier-controlled dwell-time effects in the  single-channel limit. In the semiclassical regime ($N=10$), $\mathcal{Q}^{\mathrm{QIC}} \rightarrow 0$ for both symmetry classes, in full agreement with the behavior previously observed in the Wigner–Dyson and chiral ensembles, Fig.~\ref{FigQ}(b).

For the AZ classes with ideal contacts, $Q^{\mathrm{QIC}}/Q$ reaches
$51.05\%$ of the total $\mathcal{Q}_{\mathrm{R}}$ in the DIII class and $16.84\%$ in the
CI class (see Table~\ref{table_II}). For the $\mathcal{Q}_{\mathrm{MS}}$ in the same regime,
the QIC contributions are $26.87\%$ and $6.82\%$ for DIII and CI,
respectively, confirming that the $\mathcal{Q}_{\mathrm{R}}$ is considerably more sensitive
to quantum interference from a symmetry-class perspective. In the opaque
regime (see Table~\ref{table_II}), the hierarchy is reversed: the QIC
contribution to the $\mathcal{Q}_{\mathrm{MS}}$ reaches $37.32\%$ (DIII) and $12.06\%$ (CI),
while for the $\mathcal{Q}_{\mathrm{R}}$ it is only $3.75\%$ and $0.51\%$, respectively. Furthermore, increasing the number of channels from $N=1$ to $N=10$
in the opaque regime amplifies the TRS-driven QIC contribution to the
$\mathcal{Q}_{\mathrm{R}}$ in Andreev billiards from $3.75\%$ to $16.25\%$. In contrast,
the chirality-driven contribution to the $\mathcal{Q}_{\mathrm{MS}}$ in Dirac billiards
decreases from $53.94\%$ to $22.35\%$. Thus, the two symmetry classes
display qualitatively different channel-number dependences in the
opaque regime.

\section{Conclusion}
\label{sec:conclusion}

We have investigated charge transport and thermodynamic uncertainty relations in open chaotic Andreev billiards across the Altland--Zirnbauer symmetry classes, covering the crossover from the extreme quantum ($N=1$) to the semiclassical ($N=10$) regime and from ideal to opaque contacts. At the transport level, we found that conductance and shot-noise exhibit distinct symmetry-dependent quantum-interference corrections. In particular, varying the barrier transparency drives sign changes in these corrections, revealing a crossover between weak-localization and weak-antilocalization behavior. While the conductance approaches an almost Ohmic behavior in the multichannel regime, interference signatures remain visible in the shot-noise, showing that fluctuations retain symmetry information beyond the average transport response.

For the TUR, the condition $\mathcal{Q}\geq2$ is satisfied across all symmetry classes and regimes investigated. For ideal contacts, $\mathcal{Q}_\text{R}$ provides a sensitive fingerprint of the full Altland--Zirnbauer symmetry class. In contrast to Schr\"odinger and Dirac billiards, however, the quantum-interference correction is not the dominant contribution to the TUR in Andreev billiards. This originates from the Andreev-reflection-induced enhancement of the semiclassical background, which reduces the relative weight of the interference contribution. In the opaque regime, the sensitivity is instead enhanced in $\mathcal{Q}_\text{MS}$, demonstrating that barrier transparency controls which TUR observable most clearly resolves the underlying symmetry.

These results establish a direct connection between symmetry-dependent signatures in conductance and noise and their manifestation in thermodynamic precision bounds. Natural extensions include continuous symmetry crossovers induced by magnetic fields, which would allow the evolution of the TRS-driven interference contribution to the TUR to be followed directly, as well as finite-temperature and finite-energy regimes. The inclusion of dephasing and interaction effects would further clarify the robustness of these universal symmetry fingerprints beyond the fully coherent regime considered here.

\begin{acknowledgments}
IRACL, TJAM, and MMS acknowledge financial support from INCT-SpinNanoMag (Grant No. 168483/2023-8) and CNPq. IRACL also acknowledges financial support from FAPESP (Grant No. 2025/25006-9). ARR acknowledges financial support from FAPESP (Grants No. 2026/03311-7, No. 2025/25217-0, No. 2025/24203-5, No. 2023/09820-2, and No. 2021/14375-0). DB acknowledges financial support from CNPq (Grants No. 402830/2023-7 and No. 303469/2019-6).

\end{acknowledgments}


\bibliographystyle{apsrev4-2}
\bibliography{referencias}


\end{document}